\documentclass[journal,12pt,onecolumn,draftclsnofoot,]{IEEEtran}

\usepackage{type1cm}
\usepackage{lettrine}
\usepackage{graphicx}
\usepackage{array}
\usepackage{cite}
\usepackage{float}
\usepackage{mathtools}
\usepackage{amssymb}
\usepackage{xcolor}
\usepackage[utf8]{inputenc}
\DeclareUnicodeCharacter{0301}{\'{e}}
\DeclareUnicodeCharacter{0308}{\"{o}}
\usepackage[T1]{fontenc}
\usepackage{amsmath}
\ifCLASSINFOpdf
\else
\fi
\begin{document}
%
\title{Orthogonal Frequency Division Multiplexing With Subcarrier Power Modulation\\ for Doubling the Spectral Efficiency of 6G and Beyond Networks}
%
%
%

\author{Jehad M. Hamamreh, Abdulwahab Hajar, and Mohamedou Abewa%
     \thanks{J. M. Hamamreh, A. Hajar and M. Abewa are with WISLAB-TELENG for Wireless Research at the department of Electrical-Electronics Engineering, Antalya  Bilim University, Antalya, Turkey. Corresponding author: J. M. Hamamreh (email: jehad.hamamreh@antalya.edu.tr // web: https://sites.google.com/view/wislab).}
     \thanks{This work was supported in part by the Scientific and Technological Research Council of Turkey (TUBITAK), under project grant No. 119E408.} 
      \thanks {The matlab simulation codes used to generate the results in this paper can be found at https://researcherstore.com/Simulation-Codes/OFDM-SPM}
     \thanks{Part of this research was presented at EBBT-2019 Conference \cite{SPM}.}
}

\maketitle

\begin{abstract}
With the emergence of new applications (e.g., extended reality and haptics), which require to be simultaneously served not just with low latency and sufficient reliability, but also with high spectral efficiency, future networks (i.e., 6G) should be capable of meeting this demand by introducing new effective transmission designs. Motivated by this, a novel modulation technique termed as orthogonal frequency division multiplexing with subcarrier power modulation (OFDM-SPM) is proposed for providing highly spectral-efficient data transmission with low-latency and less-complexity for future 6G wireless communication systems. OFDM-SPM utilizes the power of subcarriers in OFDM blocks as a third dimension to convey extra information bits while reducing both complexity and latency compared to conventional schemes. In this paper, the concept of OFDM-SPM is introduced and its validity as a future adopted modulation technique is investigated over a wireless multipath Rayleigh fading channel. The proposed system structure is explained, an analytical expression of the bit error rate (BER) is derived, and numerical simulations of BER and throughput performances of OFDM-SPM are carried out. OFDM-SPM is found to greatly enhance the spectral efficiency where it is capable of doubling it. Additionally, OFDM-SPM introduces negligible complexity to the system, does not exhibit error propagation, reduces the transmission delay, and decreases the transmission power by half. 
\end{abstract}


%
\IEEEpeerreviewmaketitle

%
%
%
%
\section{Introduction}
\lettrine{T}{he} technological era we live in is continuously witnessing several technological breakthroughs. Many of these upcoming technologies, however, demand requirements which current-day 4G and 4.5G systems fail to provide. As such, the next generation of 5G communication systems has been proposed. At the forefront of 5G communication and beyond systems, one of the core characteristics aimed to be improved is the data rates it can provide and the spectral efficiency it can achieve \cite{Ankarali2017}. \textcolor{black}{On the other hand, future 6G systems are expected to serve a new set of applications such as haptics, telemedicine, brain-machine interfaces, mixed reality, virtual reality, augmented reality,  etc. as pointed out in \cite{Saad6GVision}, \cite{6G-ZTE}, and some other recent studies. These services require the simultaneous achievement of 1) high spectral efficiency (high data rates), 2) low latency with good reliability, and 3) low complexity}.\footnote{The matlab simulation codes used to generate the results in this paper can be found at https://researcherstore.com/Simulation-Codes/OFDM-SPM}. 

\par Aiming to increase data rates, different approaches have been considered in the literature. Millimeter-wave transmission is one of the approaches that has been proposed to increase data rates by enabling the utilization of a larger portion of the frequency spectrum \cite{5GTech}. However, whereas mm-wave transmission provides higher data rates, its deployment has been hindered by its propagation characteristics as mm-wave communications are highly sensitive to external variables (e.g., blockage and absorption) and can only propagate for short distances. Furthermore, cell densification \cite{5GTech}, which is a rather conventional approach to improve the data rate in a given area, is also a possible way that can be used in scenarios of high traffic demands. This approach, however, incurs high costs of site rents and installation, requires time, and the effect of this is certainly only bounded to a certain area. Considered highly promising, other Multiplexing techniques besides Orthogonal Frequency Division Multiplexing (OFDM) have been introduced. These methods improve the spectral efficiency collectively within the area they are utilized, rather than improving the spectral efficiency exhibited per user. To this regard, various methods have been proposed, the most significant being Non-Orthogonal Multiple Access (NOMA) and Massive Multiple-Input-Multiple-Output (MIMO) \cite{5GTech, NOMA}. Despite achieving gains in terms of spectral efficiency, these techniques also introduce their specific drawbacks. This is so because such techniques are solely focused on improving the area spectral efficiency (i.e., number of users that can be served in an area) rather than improving the spectral efficiency per device or user. Besides, such approaches result in large processing latency, an increase in the digital signal processing complexity of the transceiver design, and low energy utilization efficiency \cite{MIMO}. OFDM based NOMA systems have also been proposed and found to achieve reasonable gains in spectral efficiency \cite{OFDMNOMA}. Another scheme capable of attaining great gains in spectral efficiency of wireless communication systems is In-band Full Duplex, which has also been proposed and studied extensively in the literature. This scheme, however, suffers from the self-interference problem, which it inherently introduces and the complex transceiver structure it requires, in addition to being incompatible with current wireless devices and standards \cite{inband}.
\par Unlike the aforementioned ways that were mostly concentrated on increasing the total system data rate or throughput by enhancing the area spectral efficiency, other approaches have opted to enhance the spectral efficiency per device/user by modifying and improving the OFDM wavefrom, which remains the backbone technology of current wireless standards (e.g., 4G-LTE, WiFi, WiGig, LiFi, DVB, etc.) as well as upcoming 5G systems. To this end, various novel techniques have appeared in the literature. For example, the authors in \cite{CPless} devised the use of a method which eliminates the need of the cyclic prefix (CP) between OFDM symbols, introducing what is know as CP-less OFDM, thus achieving gains in spectral efficiency and other performance metrics. However, an alternative approach which has attracted greater attention in the literature has been to pair the conventional OFDM waveform with an additional modulation technique that can create another dimension for conveying extra data per OFDM symbol \cite{final}. Schemes such as spatial modulation OFDM (SM-OFDM), subcarrier-index modulation OFDM (SIM-OFDM), OFDM with index modulation (OFDM-IM) and OFDM with subcarrier number modulation (OFDM-SNM) have been reported in the literature \cite{final}. A comprehensive comparative study between many of these modulation techniques in terms of various performance metrics in addition to the working principles of each proposed technique has been given and explained in \cite{final}.

\par These proposed techniques, in addition to utilizing the conventional amplitude and phase of a symbol, introduce a third data-carrying dimension to transmit and receive more data bits per OFDM block. For example, SM-OFDM \cite{SM} has utilized the indices of transmitting antennas as a third data-carrying dimension. SIM-OFDM \cite{SIM}, has instead used the indices of the active subcarriers in an OFDM block to establish a third dimension which conveys data. OFDM-IM \cite{IM}, similar to SIM-OFDM, also focuses on utilizing the indices of the active subcarriers to carry and interpret extra data bits.  OFDM-SNM \cite{SNM} furthermore, has utilized the number rather than the indices of active subcarriers of smaller OFDM sub-blocks to convey additional data bits. Although achieving some gains in spectral efficiency, many of these techniques have been impeded by certain inherent characteristics whether it is in the transceiver complexity or the scalability of the scheme. For example, it has been reported that subcarrier index dependant schemes such as OFDM-SIM and OFDM-IM provide gains mainly at low transmission rates where the benefits of these schemes incur for high transmission rates erode \cite{SIMA}. This has further been confirmed by \cite{BIGG}. Additionally, it is important to note that in some schemes such as OFDM-IM and OFDM-SNM, the deactivation of a certain number of subcarriers of the OFDM blocks or sub-blocks is an inherent requirement to establish the third data-carrying dimension and maintain its correct functionality. Consequently, this leads to less symbols being transmitted and conceptually leads to a decrease in the data rate of these systems when compared to schemes which utilize all the subcarriers of the OFDM block. This loss in data rate, however, can often be partially compensated for by the extra data bits conveyed by the additional dimension these techniques introduce, or other auxiliary methods devised by follow-up research.

 For instance, in the case of Multiple Mode (MM)-OFDM-IM \cite{MM} and Dual Mode (DM)-OFDM with Index modulation \cite{DM}, which build upon OFDM-IM, all subcarriers in the OFDM symbol are  utilized for carrying data symbols, thus yielding better spectral efficiency than plain OFDM-IM. Follow up research has led to the extension of some of the concepts of these schemes in different domains including time, frequency, space, code, etc. In \cite{IM5G}, a comprehensive survey is provided on various techniques, which have utilized index modulation in different domains including the frequency\cite{IM}, space\cite{SM}, space-time\cite{BOOK}, and space-frequency domains\cite{GSIM}, where a comparison for these scheme in terms of energy and spectral efficiency is provided. In addition to these, \cite{AIM} has introduced a number of schemes which build upon OFDM-IM and proposed algorithms to simultaneously improve both spectral efficiency and physical layer security \cite{physecsurvey,Hamamreh2018ARQ} of the system, resulting in schemes such as OFDM with Adaptive Index Modulation and Fixed Constellation Modulation (OFDM-AIM-FCM) and others. Enhanced OFDM-SNM \cite{ESNM} has also been proposed as an extension of OFDM-SNM with enhanced reliability.

Although the aforementioned OFDM-based methods can somehow improve the spectral efficiency/data rates to some extent, but this improvement is usually incremental and unfortunately comes at the cost of significant increase in complexity, delay, and overhead.
\textcolor{black}{However, with the expected emergence of new types of applications and services including augmented reality (AR), virtual reality (VR), mixed reality (MR), haptics, brain-machine interfaces, super-high definition video streaming, and real-time gaming, there will be a new set of requirements that need to be met simultaneously. These requirements are basically to provide super high throughput per device while maintaining low-latency, good reliability, and less-complexity. Such requirements are deemed beyond 5G capabilities as they do not belong to any of the three main use cases that 5G is expected to support, which are mMTC, URLLC, and eMBB \cite{6G-ZTE}. Since 5G cannot meet such requirements (see Appendix A for more technical details), future 6G and beyond wireless technologies are indeed required to address such challenges\cite{6G-ZTE, 6G-aca, 6G-aca2, Saad6GVision}}.

\textcolor{black}{To meet this demand, in this paper, a novel, low-complexity, and low-latency modulation scheme, which is capable of adding a third data-carrying dimension to double the spectral efficiency per device in future 6G and beyond networks, is developed and proposed. By the manipulation of the power of the OFDM subcarriers as an extra degree of freedom, this scheme utilizes all the subcarriers of a given OFDM block to convey extra data bits while maintaining the transmission of the modulated symbols carried by the subcarriers}\footnote{While manipulation of transmit power levels had been utilized before to convey additional data that can help eliminate the necessity for channel estimation, thus attaining reasonable gains such as in \cite{PDCE}, the proposed OFDM-SPM scheme, which utilizes the power of subcarriers in the OFDM structure, is a completely very different technique that offers unprecedented gains in terms of spectral efficiency with low-complexity.}. The main contributions of this paper can be summarized as follows:
\begin{itemize}
    \item The proposition and introduction of a novel modulation scheme known as OFDM-SPM, and the design of its transceiver structure and working principles.
    \item The performance analysis of this novel scheme in terms of power and spectral efficiency is provided. Additionally, an analytical analysis of the theoretical BER of the system is derived and given in a handy closed-form expression.
    \item The formulation of two different policies of OFDM-SPM, where one allows power saving, whereas the other improves BER performance using power reallocation. Furthermore, finding the optimal values of the power levels which provide the optimal BER performance for both polices is conducted.
    \item Providing comprehensive numerical simulation results for OFDM-SPM in terms of BER and throughput performances under different use cases and for both the power saving and power reallocation policies.
\end{itemize}

\par Comparing the proposed OFDM-SPM\footnote{It has been reported in the literature that in the domain of optical communications, power has been utilized alongside index modulation as an extra degree of freedom to send extra data bits \cite{SIPM}. This, however, yielded only minor gains in the spectral efficiency when compared to the proposed wireless-tailored OFDM-SPM scheme.} with conventional OFDM, OFDM-SPM has shown to offer a significant advantage over conventional OFDM in terms of spectral efficiency. Overall, OFDM-SPM is superior to conventional OFDM as it offers the following merits:
\begin{itemize}
\item The spectral gain which amounts to doubling the spectral efficiency of the system \textbf{using only one sinusoidal carrier} unlike conventional modulation schemes (e.g., QPSK, M-QAM, M-PSK) that have to use two orthogonal carriers (sin and cosine) to improve spectral efficiency. From another standpoint, OFDM-SPM combined with binary phase-shift keying (BPSK) symbol modulation can transfer as much data as conventional OFDM with BPSK \textbf{using only half the number of subcarriers that conventional OFDM requires}. 
\item As the number of subcarriers used by OFDM-SPM is half the number of those used in conventional OFDM. OFDM-SPM with BPSK also reduces the transmission delay for the same amount of throughput, since a fewer number of subcarriers translates to fewer resources in the time domain as well. Furthermore, it is capable of reducing complexity as it can use half of the IFFT size that OFDM would require for achieving the same throughput. Also, OFDM-SPM detection process is another source of reducing complexity because it uses simple threshed-based detectors  unlike those schemes that depend on using maximum likelihood detectors which involve high complexity \cite{IM5G}, especially for large mapping tables. 
\item OFDM-SPM offers flexibility in the option of reducing the transmission power by half while maintaining the spectral gain, at the expense of some degradation in the BER or reallocating the saved power for an improvement in the system BER.
\end{itemize}

\par The aforementioned merits of OFDM-SPM clearly characterize it as being a system which requires low complexity as the complexity it adds to the system transceiver structure is minor. Additionally, it is characterized by small-time delays and offers high spectral efficiency. This makes OFDM-SPM an ideal fit for certain applications in the IoT domain, where low latency, low complexity, and high efficiency are all required. 

\par The remaining sections of this paper are organized as follows. Section II explains the system model of the proposed OFDM-SPM scheme. In Section III, the performance analysis is carried out. Section IV provides the system's performance demonstration and elaborates on it. Finally, section V presents the conclusion and future possible works related to the proposed scheme.

\section{OFDM-SPM: System Model}
In this section, we explain and illustrate in detail the transmitter and receiver designs of the proposed OFDM-SPM scheme.
\subsection{The Transmitter Design}
OFDM-SPM mainly aims to transmit more bits per subcarrier by manipulating the power of the subcarriers in an OFDM block in addition to those bits that are usually transmitted by conventional modulation schemes such as BPSK, M-PSK, etc. The general architecture of the OFDM-SPM transmitter is depicted in Fig. 1.
\begin{figure}[h]
\centering
\includegraphics[width = 10cm, height = 6cm]{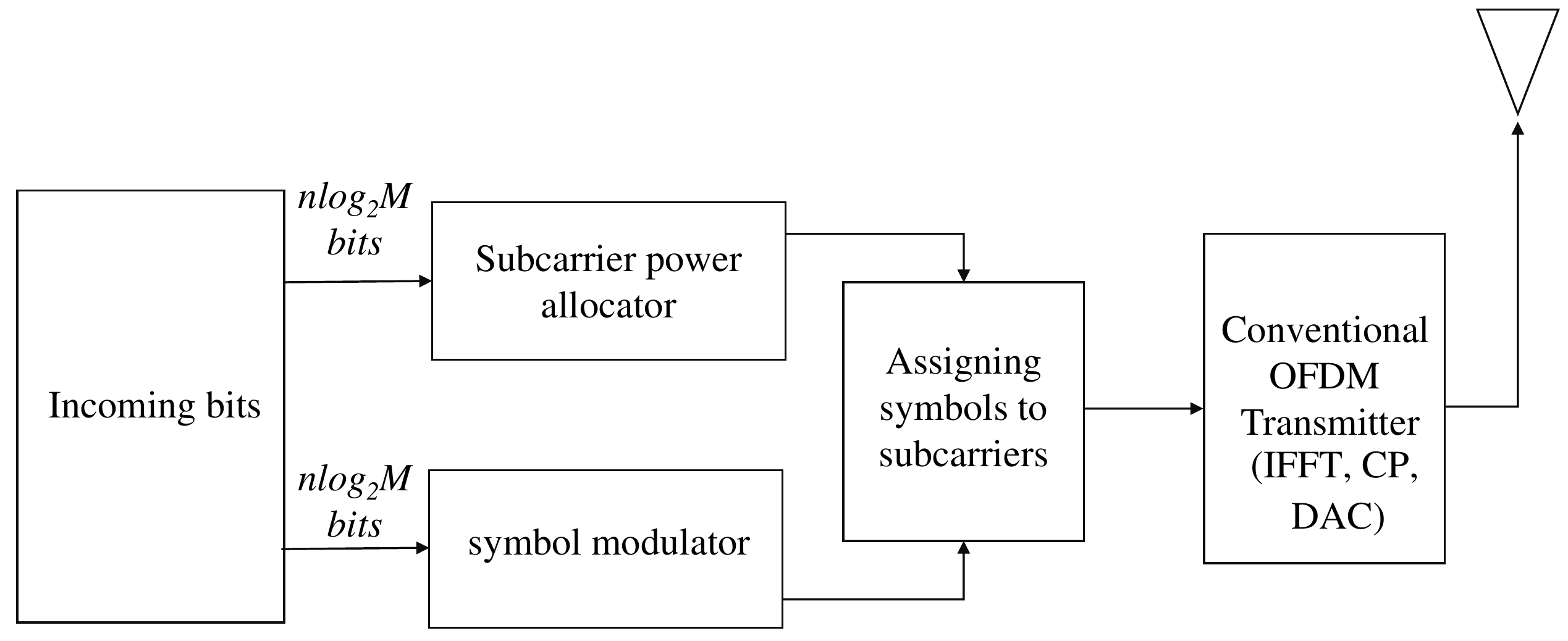}
\caption{Transmitter structure of OFDM-SPM.}
\end{figure}
\par Unlike the conventional OFDM, OFDM-SPM splits the serial input bit stream of length $2n\log_{2}M$ into two sub-streams of $n\log_{2}M$ bits, where ${n}$ is the number of subcarriers in the OFDM block used to carry data and $M$ represents the modulation order, which in the case of BPSK is 2. As such, $\log_{2}M$ is the number of bits per symbol according to the symbol modulation scheme used.

In this study, we consider pairing the concept of OFDM-SPM with BPSK symbol modulation. The reasons behind this selection are given as follows: 
\begin{itemize}
    \item To simplify the explanation and interpretation of the proposed OFDM-SPM concept so that it can be understood clearly  and easily.
     \item The fact that BPSK is the least complex coherent modulation scheme as it uses only one carrier, resulting in avoiding the I/Q problem that exists in other two-carriers based modulation schemes such as M-QAM
     \item Last but not least, BPSK is the most highly reliable modulation scheme where it has the lowest error rates among all other conventional modulation schemes.  
\end{itemize}

These last two points mentioned above suite very well the 6G use-case (i.e., services that require low complexity, good reliability, and high spectral efficiency) that we try to satisfy its requirements by the proposed OFDM-SPM. Nevertheless, the proposed OFDM-SPM concept can be integrated and paired smoothly with other modulation orders such as QPSK, M-PSK and M-QAM. This integration process is left for future studies to investigate and quantify the effect of high modulation orders on the performance of OFDM-SPM. 

As can be seen from Fig. 1, one of the substreams determines the power levels of the subcarriers, where the \textit{i\textsuperscript{th}} bit determines the power level of the \textit{i\textsuperscript{th}} subcarrier utilized to carry data. A '1' bit corresponds to setting the power of the respective subcarrier to high and a '0' bit to low. The second sub-stream of bits is modulated using the regular BPSK modulation scheme. The BPSK symbols are then assigned to their respective subcarriers. Finally, the symbols go through the remaining steps of the conventional OFDM transmission process, including inverse fast Fourier transform (IFFT), normal cyclic prefix (CP) addition, and digital-to-analog conversion (DAC). \\Thus, OFDM-SPM results in 4 different constellation points as shown in Fig. 2.

\begin{figure}[H]
\centering
\includegraphics[scale = 0.35]{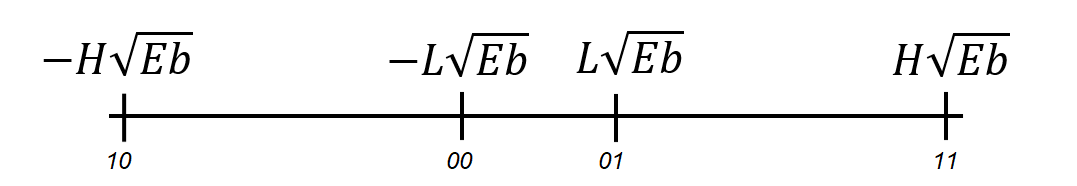}
\caption{Constellation points of OFDM-SPM with BPSK.}
\end{figure}

In Fig. 2, ‘00’ refers to a ‘0’ modulated by BPSK, which is carried by a low power subcarrier that represents another '0' as well. Similarly, ‘11’ refers to a ‘1’ modulated by BPSK, which is carried by a high power subcarrier that represents '1' as well. Also, ‘01’ corresponds to a ‘1’ modulated by BPSK, which is carried by a low power subcarrier that represent '0', and vice versa as visualized in Fig. 3. Furthermore, \textit{L} and \textit{H} denote the low and high power levels of the subcarriers, respectively, whereas \textit{E\textsubscript{b}} denotes the energy per bit. More specifically, \textit{L} and  \textit{H} are factors which determine the power of a subcarrier relative to the power given to a BPSK symbol, which is generally normalized to unity, such that a subcarrier with \textit{H} power indicates a subcarrier with \textit{H} times the power of a BPSK symbol, and similarly for \textit{L}.

As the assigned \textit{H} and \textit{L} power factors affect the overall bit error rate of the scheme, these values were chosen optimally in order to minimize this error. This optimization was done under a constraint that ensures the average energy of an OFDM subcarrier in OFDM-SPM can not exceed that of a subcarrier in conventional OFDM using BPSK symbol modulation. This allows a fair comparison between both schemes and ascertains that OFDM-SPM achieves a great gain in spectral efficiency without requiring additional power. The following example will demonstrate that OFDM-SPM, in addition to its spectral efficiency gain, requires less power than conventional OFDM with BPSK to transmit the same amount of data bits.

\begin{figure}[H]
\centering
\includegraphics[width = 9cm, height = 13cm]{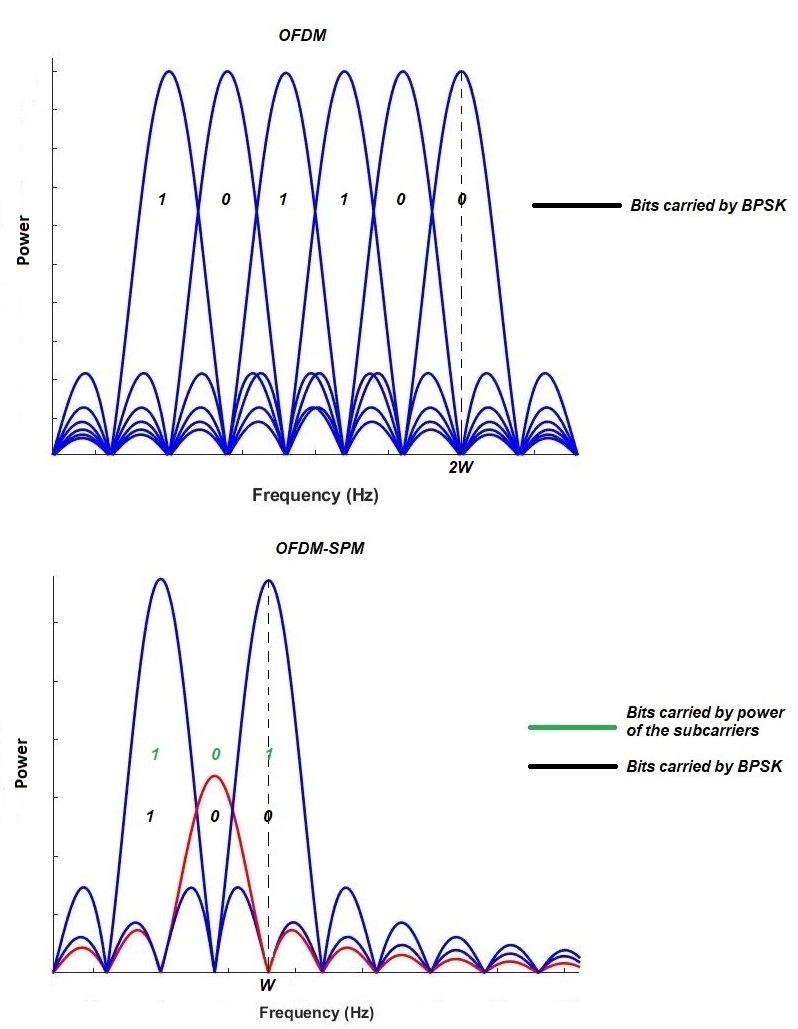}
\caption{Comparing OFDM-SPM with OFDM, where half the bandwidth and half the power are saved by OFDM-SPM.}
\end{figure}

\par $\textbf{Example:}$ To clearly illustrate the merits that OFDM-SPM offers, let us display a case where both conventional OFDM and OFDM-SPM can be used. Assuming 104 bits were required to be transferred. Conventional OFDM with BPSK, since it carries only one bit per subcarrier, would require a total of 104 subcarriers, each requiring bandwidth \textit{W} and power \textit{P}. As such the total resources used by conventional OFDM would be \textit{104W} and \textit{104P} in terms of bandwidth and power, respectively. OFDM-SPM on the other hand, would require only 52 subcarriers, as each subcarrier carries two bits, resulting in a total resource usage of \textit{52W}, and \textit{52P}.

\par As illustrated in Fig. 3, OFDM-SPM doubles the spectral efficiency where conventional OFDM as per the figure uses \textit{2W} bandwidth, whereas OFDM-SPM only requires \textit{W} bandwidth for the same number of bits transmitted. Additionally, OFDM-SPM reduces the total power usage by half as a consequence of using only half the number of subcarriers used by conventional OFDM. This saved power provides OFDM-SPM with flexibility, since this power can either be saved or reallocated according to the requirements of the application. Applications requiring low complexity and low power can benefit from this merit of OFDM-SPM, thus presenting OFDM-SPM as a favorable modulation scheme for applications such as IoT. For applications requiring a more highly performing BER, the saved power can be reallocated, which would result in an enhanced BER performance. 
\par The optimal values of the high and low power levels, which minimize the BER of the scheme are found by means of a successive process of exhaustive trial and error experiments. The power levels in the case of power saving and power reallocation policies \cite{SIM} are defined according to Eq. (1) and (2)\footnote{A further detailed explanation on Eq. (1) and (2), and how the optimal \textit{L} and \textit{H} power levels were found are given in Appendix B and Appendix C, respectively, for the convenience of the reader.}, respectively.\begin{equation} L^2 + H^2 = 2E_{b}\end{equation}\begin{equation} L^2 + H^2 = 4E_{b}\end{equation}. 

In the case of power saving for example, Eq. (1) is followed to determine the values of \textit{H} and \textit{L}. By setting the value of \textit{H} to an arbitrary value, the corresponding value of \textit{L} can be found as follows:

\begin{equation} L = \sqrt{2E_b - H^2} \end{equation}

\par The optimal values of \textit{L} and \textit{H} in the case of power reallocation were found in a similar manner, but using Eq. (2) instead of Eq.  (1).


Various simulations were run for varying values of \textit{L} and \textit{H}, the optimal values resulting in the minimum BER of the scheme were found as 
\textit{L = 0.4213} and \textit{H = 1.35} when power is saved rather than reallocated, and \textit{L = 0.5668} and \textit{H = 1.918} when power is reallocated.

\subsection{Channel Model}
\par The channel is assumed to be a slowly varying, Rayleigh multi-path fading wireless channel with $K$ number of exponentially decaying taps, denoted by $\boldsymbol {h}=[h_{0}, h_{1}, \cdots, h_{(k-1)}]$ \cite{OFDM-SIS, SNM}. As such the received symbols in time domain are given as:
\begin{equation}
    \boldsymbol{y} = \boldsymbol{x} \circledast \boldsymbol{h} + \boldsymbol{n},
\end{equation}
where $\circledast$ represents the convolution operation. Also, the symbols $\boldsymbol{x}$, $\boldsymbol{y}$, $\boldsymbol{h}$ and $\boldsymbol{n}$ are vectors representing the transmitted time domain samples, received samples in the time domain, the channel impulse response, and the additive white Gaussian noise, respectively. Furthermore, the $\boldsymbol{n}$ vector can be statistically characterized by   \[
\boldsymbol{n} \sim \mathcal{N}(0,\,N_{0})\,,
    \] 
where the elements (noise samples) of $\boldsymbol{n}$ have zero mean variance equal to $N_{0}$. Additionally, the channel is slowly varying in time such that it is assumed to be constant for multiple OFDM symbols duration before it changes independently in the subsequent time intervals. Eq. (4) can also be represented in the frequency domain as well, which is more handy to deal with as can be found in  \cite{CPless} and \cite{OFDM-SIS}. 

\subsection{The Receiver Design}
OFDM-SPM has a significant merit in its receiver as its structure is very simple and adds minor complexity to the overall receiver structure of conventional OFDM (i.e., highly desirable for future wireless standards). The receiver structure is displayed in Fig. 4. Similar to conventional OFDM, the incoming data goes through many of the conventional processes of OFDM reception including ADC, CP removal, FFT, etc. At the symbol demodulation stage, however, the signal is fed into two parallel blocks for detection. The first block is responsible for non-coherently detecting the first sequence of bits, which modulated the power levels of the OFDM subcarriers. By comparing the power levels of each subcarrier to a given threshold \textit{T}, the power levels can be detected as high or low, or equally stated the bits can be detected as a '1' or a '0', respectively. Particularly, if the power of a subcarrier is greater than the threshold, a '1' is detected and vice versa. 
This threshold \textit{T} is optimally determined as the power of the midpoint value between the high and low amplitude values of the subcarriers. On the other hand, the other detection block performs conventional coherent BPSK demodulation to the symbols carried by the subcarriers.

\begin{figure}[H]
\centering
\includegraphics[width = 9cm, height = 4.5cm]{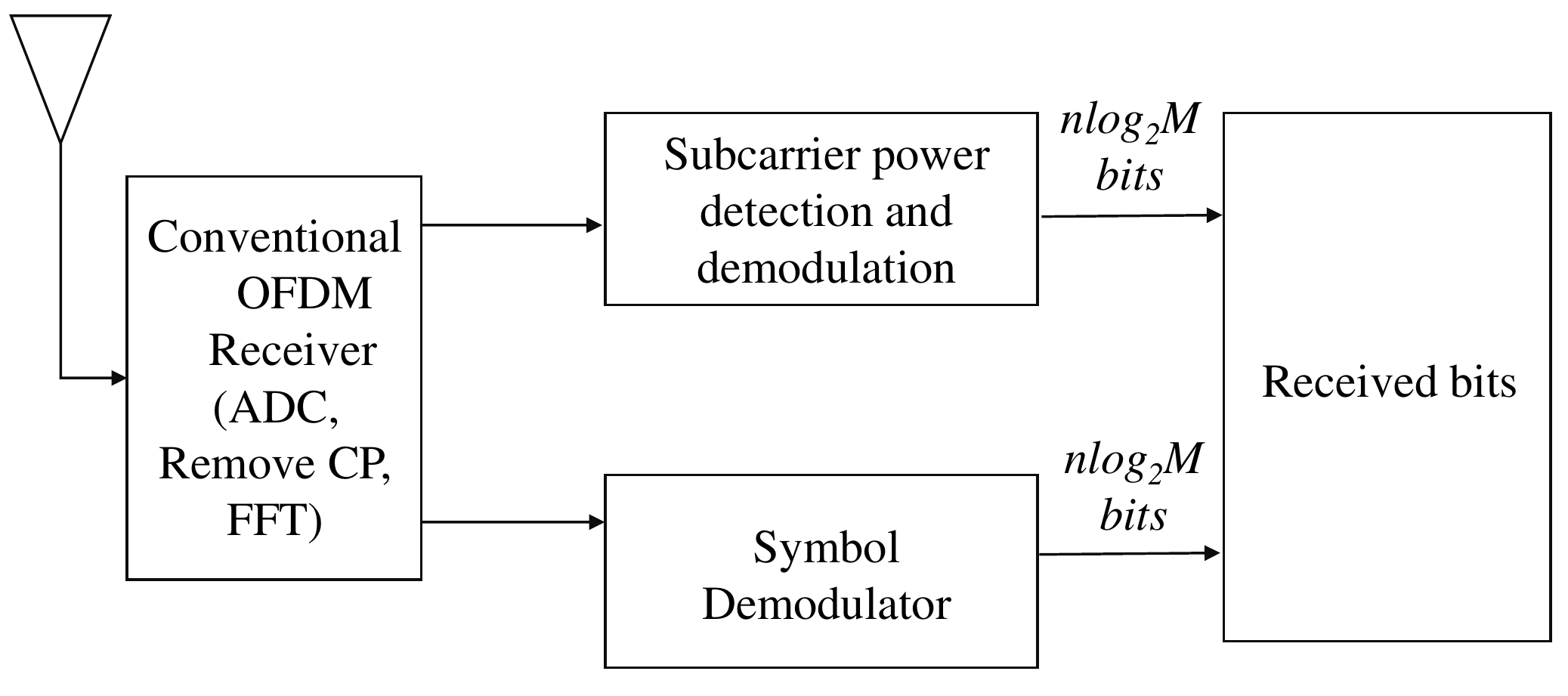}
\caption{Receiver structure of OFDM-SPM.}
\end{figure}

\par The midpoint and threshold \textit{T} used in the detection process for comparison with the power levels of the received subcarriers follow Eq. (5) and (6), as follows.

\begin{equation}
midpoint = \bigg{(}\frac{L + H}{2}\bigg{)}
\end{equation}

\begin{equation}
T = \bigg{(}\frac{L + H}{2}\bigg{)}^2
\end{equation}

It is very essential to take note of the minority of the complexity added to the system receiver structure by the non-coherent detection of the bits conveyed by the power of the subcarriers. As this detection is done by means of thresholding, which adds negligible complexity to the system. In terms of complexity, this gives OFDM-SPM an advantage over other schemes such as OFDM-SNM, OFDM-IM, SIM-OFDM, etc., which usually employ either a maximum likelihood (ML) detectors for optimum performance, or a log likelihood ratio (LLR) detector for reduced complexity \cite{final}. Both of these, however, introduce far more complexity to the receiver structure than that of OFDM-SPM. 

\section{Performance Analysis}
\subsection{Power and Spectral Efficiency}
\par Unlike other OFDM-based modulation schemes proposed in the literature for sending additional data bits by utilizing an extra dimension besides the conventional 2D signal constellation\cite{final}, in OFDM-SPM, all the subcarriers are used for sending data, where there are no inactive subcarriers and the number of active subcarriers in each OFDM block does not vary. In particular, since all the subcarriers are utilized for data transmission, measuring the superior spectral and power efficiency of OFDM-SPM is rather simple. Referring to Fig. 3, it is evident that OFDM-SPM always attains a doubling of the spectral efficiency, and a reduction of the transmission power by half.

\subsection{Bit Error Rate (BER)}
As OFDM-SPM transmits two different bit sequences, a bit can be received in error due to one of two possible reasons. Firstly, a bit can be received in error during the detection of the power levels of the subcarriers incorrectly. This is because of the combined channel effects of both the noise and the multipath fading as they can either amplify or attenuate the power of the subcarriers, leading to the detection of a high power subcarrier as a low power subcarrier and vice versa. Secondly, an error that can arise from the detection of the BPSK modulated symbols carried by the subcarriers.

\par From Fig. 2, it is evident that the power of the subcarrier assigned to a given BPSK symbol affects its probability of being detected in error. Fig. 2 shows that the constellation points with low power subcarriers exhibit a smaller minimum Euclidian distance between them causing them to be more prone to errors. Although this affects the BER performance of OFDM-SPM negatively, this is partially compensated for by the low probability of error of the constellation points with high power subcarriers which exhibit a larger Euclidian distance between them. Therefore, the theoretical BER of OFDM-SPM due to BPSK demodulation is the average of the BER expressions of its low power and high power BPSK schemes. These expressions are similar to that of BPSK in a Rayleigh fading channel, and can be found by the addition of a multiplication factor to account for the different power levels of the subcarriers. Eq. (7), (8) and (9) display the resulting BER expressions.

\begin{equation}
BER_{H} = \frac{1}{2}\bigg{(} 1 - \sqrt{\frac{H^2 \frac{E_b}{N_0}}{1 + H^2 \frac{E_b}{N_0}}}\bigg{ )}
\end{equation}

\begin{equation}
BER_{L} = \frac{1}{2}\bigg{(} 1 - \sqrt{\frac{L^2 \frac{E_b}{N_0}}{1 + L^2 \frac{E_b}{N_0}}}\bigg{ )}
\end{equation}

\begin{equation}
BER_{BPSK} = \frac {BER_{L} + BER_{H}} {2}
\end{equation}
\par Furthermore, the analytical expression of the BER resulting from the detection of the power of the subcarriers is given as per Eq. (10), (11), (12) and (13).
\begin{equation}
BER_P = A + B - C,
\end{equation}
where $A$, $B$ and $C$ are the final terms in the BER expression after the mathematical derivation and the collection of terms. $A$, $B$ and $C$ mathematical expressions are found respectively as follows:
\begin{equation}
A = \frac{1}{2}\bigg{(} 1 - \sqrt{\frac{(\frac{H - L}{2})^2 \frac{E_b}{N_0}}{1 + {(\frac{H - L}{2})}^{2} \frac{E_b}{N_0}}} \bigg{)}
\end{equation}

\begin{equation}
B = \frac{1}{4}\bigg{(} 1 - \sqrt{\frac{(\frac{H + 3L}{2})^2 \frac{E_b}{N_0}}{1 + (\frac{H + 3L}{2})^2 \frac{E_b}{N_0}}} \bigg{)}\end{equation}

\begin{equation}
C = \frac{1}{4}\bigg{(} 1 - \sqrt{\frac{(\frac{3H + L}{2})^2 \frac{E_b}{N_0}}{1 + (\frac{3H + L}{2})^2 \frac{E_b}{N_0}}} \bigg{)}\end{equation}
\par Thus, the total average BER of OFDM-SPM with BPSK is theoretically the average of  Eq. (9) and (10).
\begin{equation}
BER_{OFDM-SPM} = \frac{BER_{P} + BER_{BPSK}}{2} 
\end{equation}

The detailed derivation of this BER formula along with its main expressions' terms is given in \textbf{Appendix D}.

\par It is important to observe that the establishment of the third data-carrying dimension (i.e., power domain) by OFDM-SPM does not inherently lead to any kind of error propagation during the detection stages of the scheme. This is an additional merit for OFDM-SPM, where other OFDM-based modulation schemes such as SIM-OFDM \cite{ESIM} were rather excessively affected by the error propagation they induce. 

Furthermore, the detection processes of the BPSK bits and power bits can be seen as independent, where the detection of the subcarrier power incorrectly does not necessarily incur an error in the detection of the carried symbol. This is due to the dissimilarity between the detection processes, where the BPSK detector is coherent, whereas the power detector is non-coherent. This is so because BPSK detection measures the phase of a carried BPSK symbol, whereas the power detection involves measuring the power level of the subcarrier.

\section{Performance Demonstration}
Numerical simulations\footnote{The matlab simulation codes used to generate the results in this paper can be found at https://researcherstore.com/Simulation-Codes/OFDM-SPM} displaying the BER and throughput performance of OFDM-SPM were conducted. Table I shows the simulation parameters adopted in this study. The system was simulated in a multipath Rayleigh fading environment. The channel is slowly time-varying such that it is assumed to be constant for a block of OFDM symbols, but changes independently from one to another. This choice of the channel environment is because a multipath Rayleigh fading channel strongly mimics the characteristics of realtime wireless channels. Furthermore, OFDM-SPM has been paired with BPSK symbol modulation in order to test and introduce the idea of subcarrier power modulation clearly, while avoiding adding extra complexities that may be related to higher order symbol modulation techniques. Many other prominent techniques have been introduced and tested under identical conditions, such as in \cite{IM}, \cite{SNM}, \cite{GIM}, \cite{DM}, \cite{GDMIM}, \cite{MM} and \cite{ESNM}. The simulation results are displayed under different power allocation policies.
\begin{table}[h]
\centering
\caption{Simulation Parameters}
\begin{tabular}{ | m{18em} | m{3cm} | }
\hline
Modulation type & BPSK (\textit{M} = 2) \\
\hline
IFFT / FFT size & 64 \\
\hline
Subcarriers for data \textit{n} & 52\\
\hline
Symbols allocated for cyclic prefix & 16\\
\hline
Number of inactive subcarriers for out of band emission & 12 \\
\hline
Number of OFDM symbols & 5 $\times$ 10\textsuperscript{4} \\
\hline
Multipath channel delay samples locations & [0 3 5 6 8] \\
\hline
Multipath channel tap power profile (dBm) & [0 -8 -17 -21 -25]\\
\hline 
\end{tabular}
\end{table}

\subsection{Power Saving Policy}
By referring to Fig. 3, we can see that OFDM-SPM uses only half the number of subcarriers that conventional OFDM would require to send the same number of data bits. Thus, half the power used by conventional OFDM is unused (i.e., saved) by OFDM-SPM. In the power saving policy, this power is saved to match requirements of low power applications (e.g., IoT). This inherently results in a better power efficiency when compared to conventional OFDM.
The power levels in the simulation of this case are defined as in Eq. (1), and were found as \textit{H} = 1.35 and \textit{L} = 0.4213.

\begin{figure}[H]
\centering
\includegraphics[width = 9.5cm, height = 7.5cm]{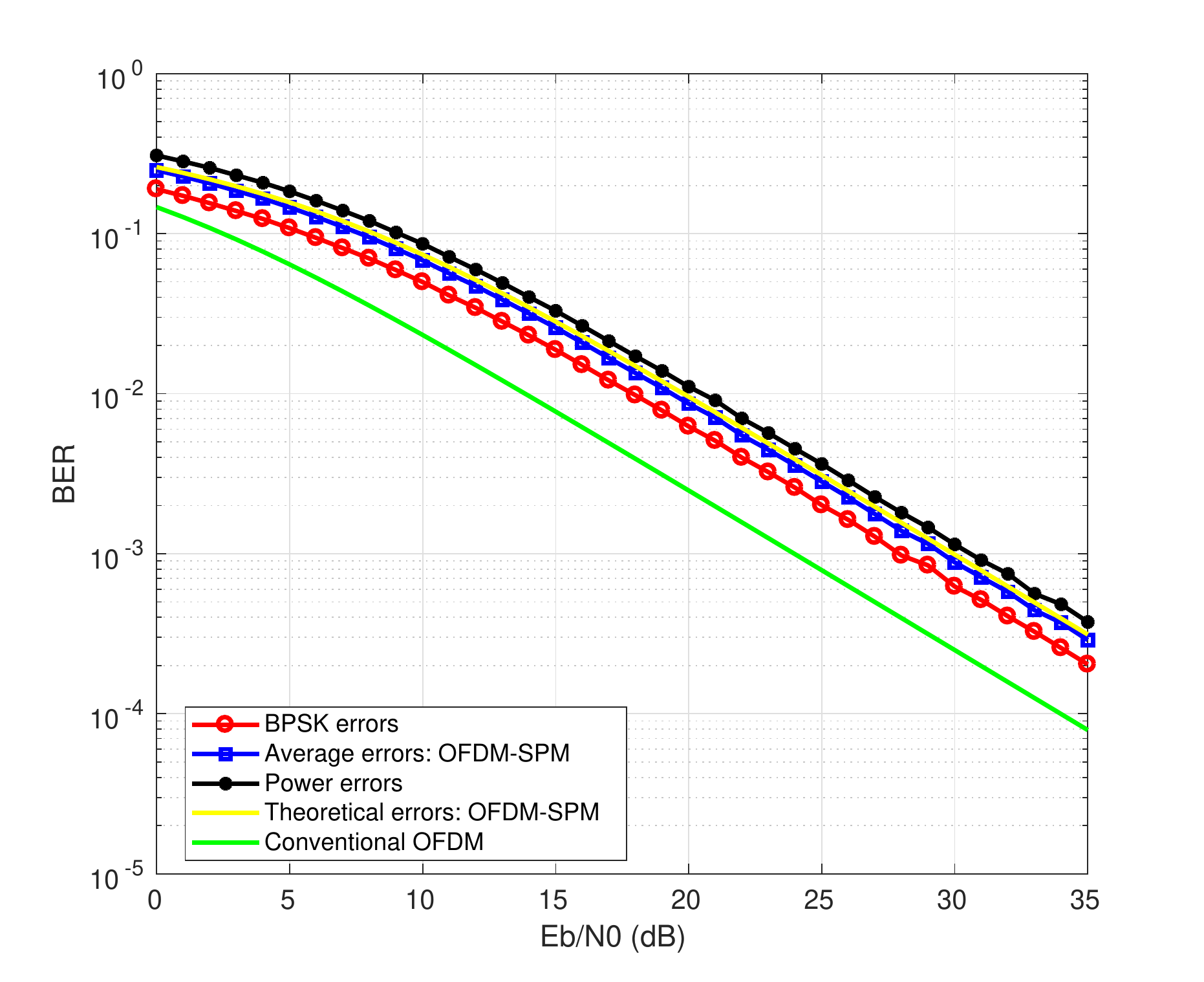}
\caption{BER of OFDM-SPM with power saving policy, where half of the transmit power is saved by OFDM-SPM.}
\end{figure}

As can be seen from the results in Fig. 5, the performance of OFDM-SPM when compared to conventional OFDM with BPSK\footnote{The reason why OFDM-SPM is compared to OFDM-BPSK (not OFDM-QPSK) is given in Appendix E.} displays some degradation in the BER, however at the expense of this degradation, it can be seen that the throughput is doubled for high values of SNR amounting to a doubling in the data rate as can be seen from Fig. 6, while reducing the transmission power by half. Furthermore, even for values as low as 20 dB a throughput of 2 (bits/s/Hz) is observed, which is a large gain in data rate. More detailed discussion on the trade-off relationship between throughput and BER in OFDM-SPM can be found in Appendix F.

\begin{figure}[H]
\centering
\includegraphics[width = 9.5cm, height = 7.5cm]{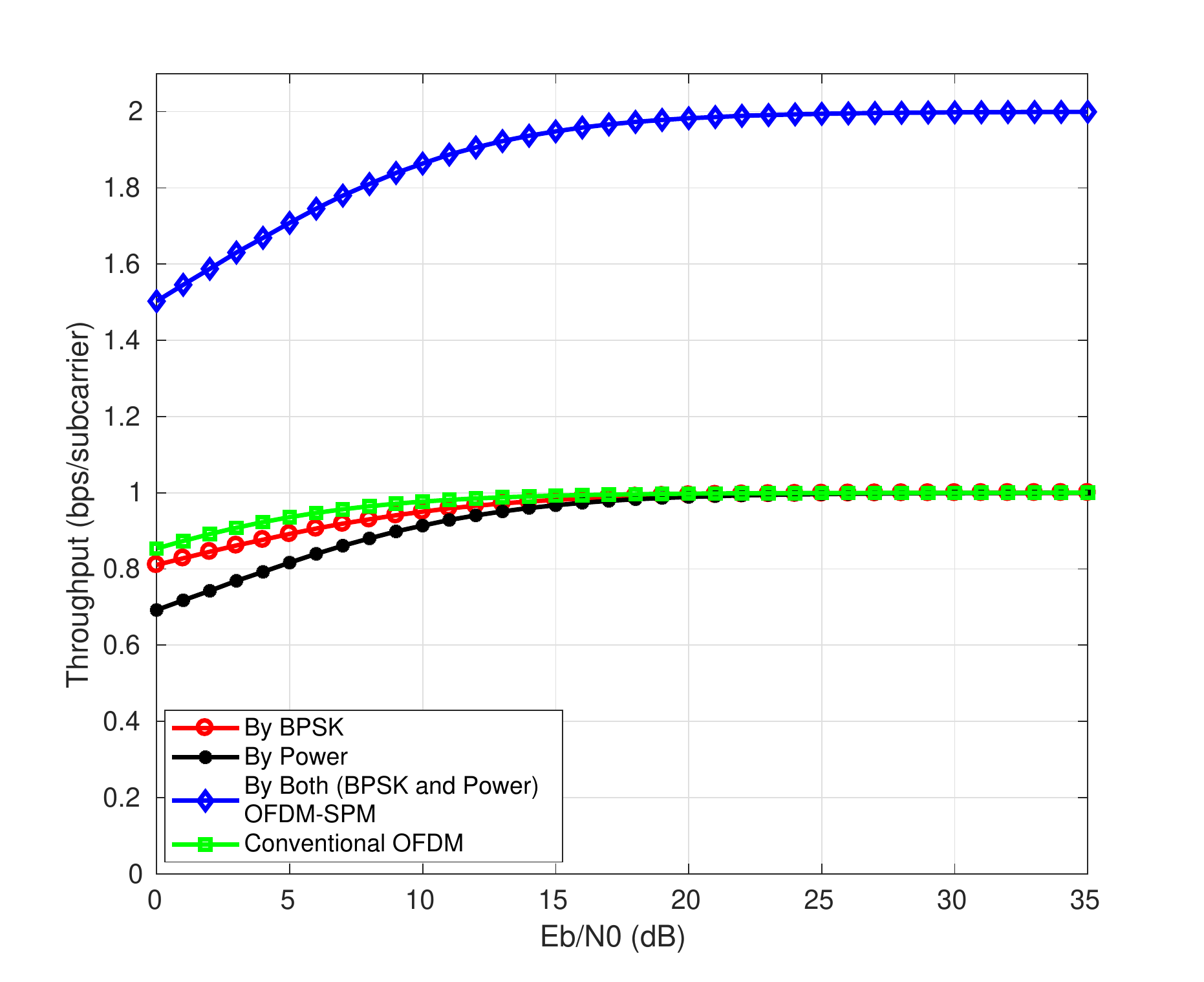}
\caption{Throughput of OFDM-SPM with power saving policy, where half of the transmit power is saved by OFDM-SPM.}
\end{figure}

\subsection{Power Reallocation Policy}

\begin{figure}[H]
\centering
\includegraphics[width = 9.5cm, height = 7.5cm]{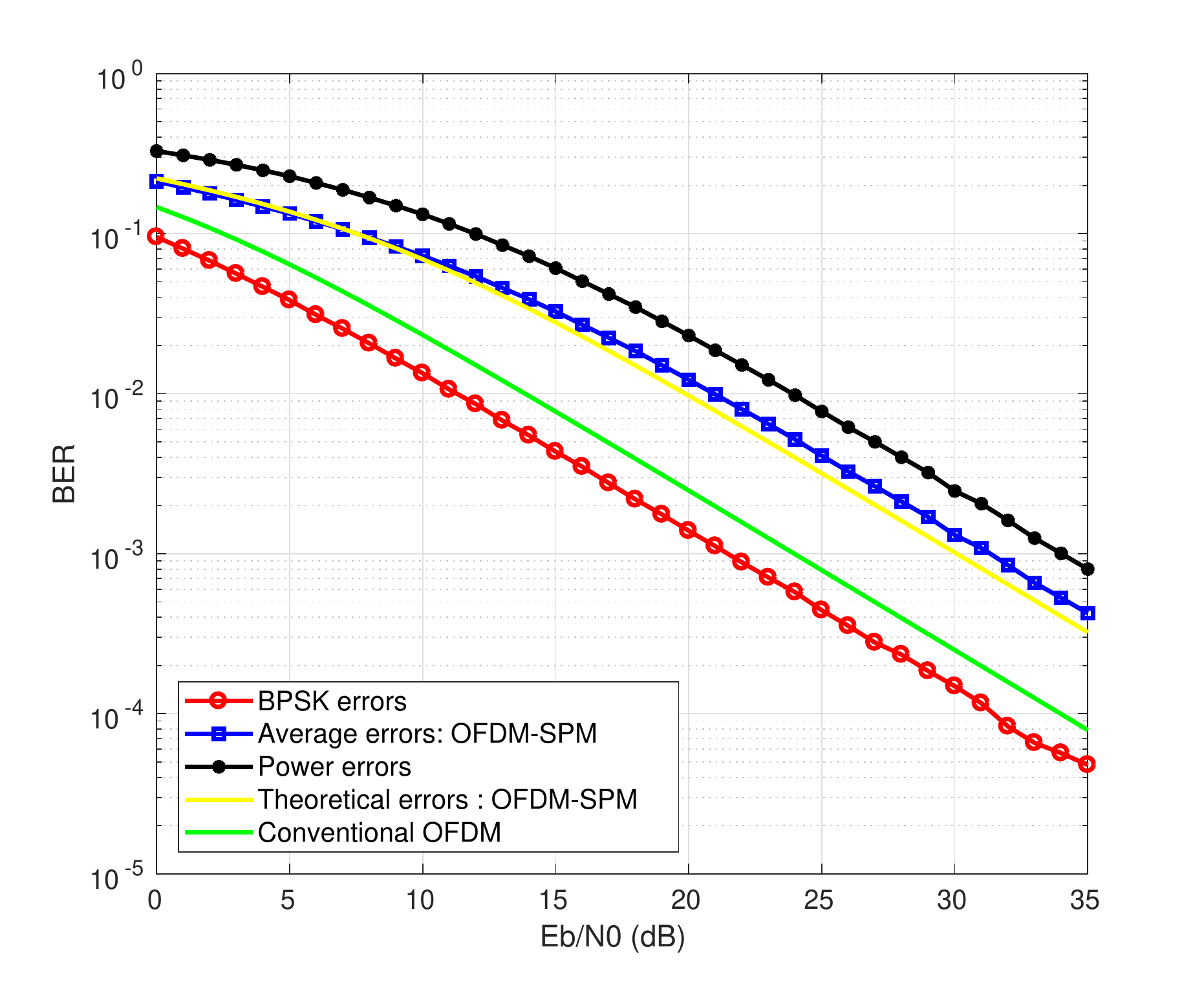}
\caption{BER of OFDM-SPM with non-optimized power reallocation policy, where the saved power is reallocated to high-power subcarriers, whereas low-power subcarriers are set with unity power levels. As can be seen, there is about 3dB improvements in the BER of the bits' stream modulated by BPSK, while having another additional bits' stream modulated by SPM.}
\end{figure}
\par The power which OFDM-SPM saves can be utilized to the scheme's advantage. As the power saving policy shows a degradation in the BER, this saved power can be reallocated to the subcarriers of the OFDM symbol, resulting in an enhanced BER. The method through which the saved power is redistributed amongst the subcarriers will ultimately affect the BER. Power reallocation is investigated for two different cases.
\subsubsection{Non-Optimized Power Reallocation Policy}

\par In this scenario, the gain of OFDM-SPM is clearly displayed. Unlike conventional OFDM, OFDM-SPM can be viewed as a transmission method that provides two streams of data. One of them being the bits carried by the BPSK symbols, and the other being the bits carried by the power of the OFDM subcarriers. In this scheme, the saved power is reallocated or reassigned such that the bit error rate performance of the bits carried by the BPSK symbols does not degrade but is rather less erroneous than the case of conventional OFDM. This is done by setting the high power level of the subcarriers to \textit{H = 1.732}, and the low power level to \textit{L  = 1}. This BPSK data stream (the red curve in Fig. 7) exhibits a bit error rate superior to that of conventional OFDM by a 2-3 dB gain as can be seen in Fig. 7. This is because the high power bits map to constellation points which are further apart than that of conventional OFDM. Furthermore, an additional data stream is provided by the bits carried by the power levels of the subcarriers. Although, this power-modulated stream exhibits a high frequency of errors (the black curve in Fig. 7), it can be seen as a mere additional benefit to the enhanced BPSK bit error rate. Additionally, this erroneous bit stream can be assigned to a user application that does not require ultra reliability such as audio streaming services.
\begin{figure}[H]
\centering
\includegraphics[width = 9.5cm, height = 7.5cm]{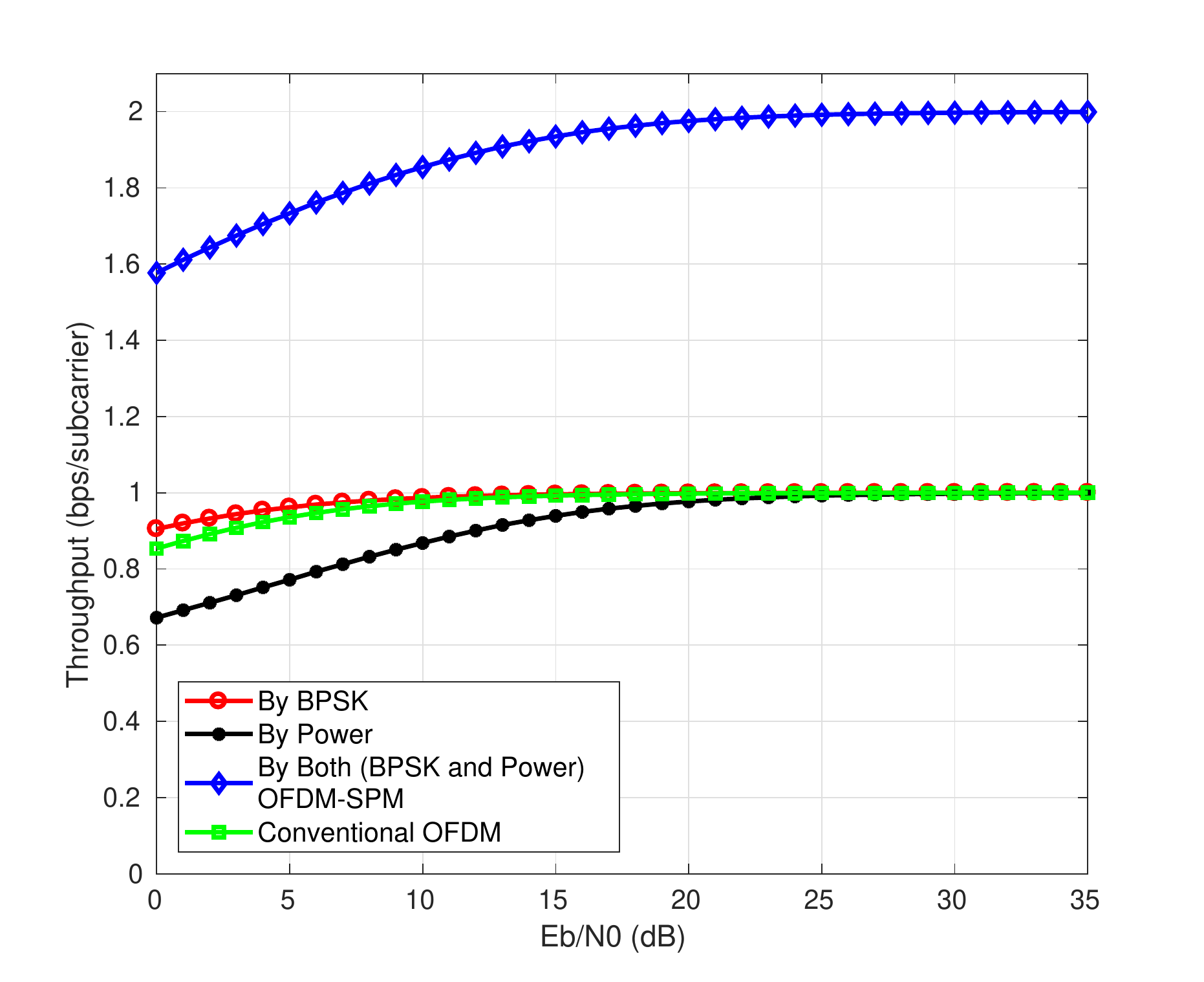}
\caption{Throughput of OFDM-SPM with non-optimized power reallocation policy, where the saved power is reallocated to high-power subcarriers, whereas low-power subcarriers are set with unity power levels.}
\end{figure}

Furthermore, the gains in terms of throughput are evident from Fig. 8, as a throughput of 1.95 (bits/s/Hz) is achieved at SNR values as low as 15 dB, and saturates at a throughput value of 2 (bits/s/Hz) at high SNR values. This basically reflects the scheme's capability in doubling the spectral efficiency of the system.

\subsubsection{Optimized Power Reallocation Policy}
In this case, the optimal power levels which provide optimal average BER for OFDM-SPM were found according to Eq. (2). Exhaustive trial and error optimization was used to find the corresponding optimal \textit{H} and \textit{L} values, which were found as \textit{H} = 1.918 and \textit{L} = 0.5668. As Fig. 9 shows, the bit error rate performance of the scheme is improved by 3 dB compared to the power saving policy presented in Fig. 5. Although a deterioration in the bit error rate is still observed when compared to conventional OFDM in a Rayleigh fading channel, the gains that OFDM-SPM offer can outweigh this slight, insignificant degradation. 

\begin{figure}[H]
\centering
\includegraphics[width = 9.5cm, height = 7.5cm]{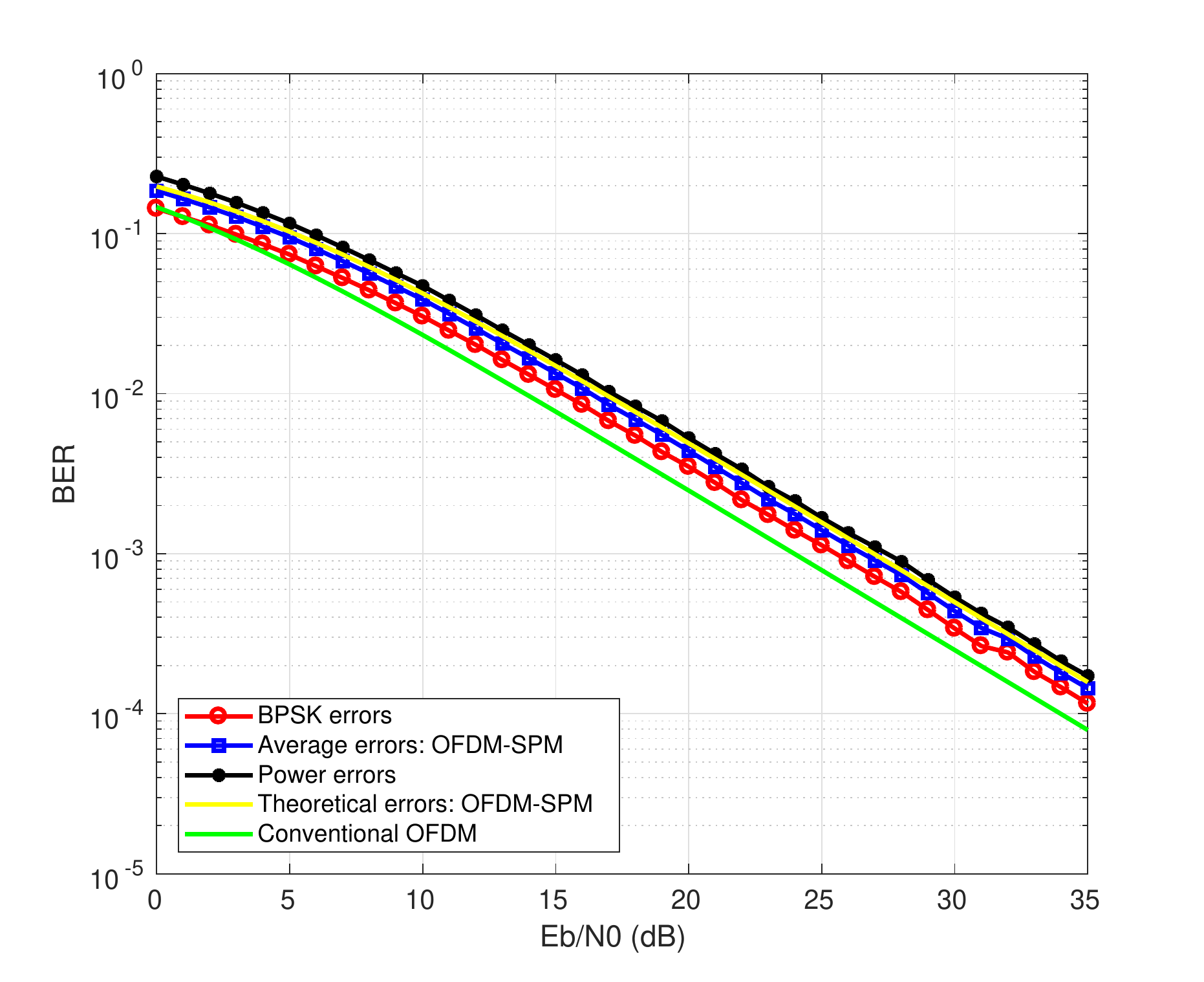}
\caption{BER of OFDM-SPM with optimized power reallocation policy, where the optimal power levels that minimize the overall average error rates are used.}
\end{figure}

Fig. 10 also displays the throughput performance of OFDM-SPM with optimized power reallocation, where for values as low as 10 dB a throughput of 1.95 (bits/s/Hz) is observed.

\begin{figure}[H]
\centering
\includegraphics[width = 9.5cm, height = 7.5cm]{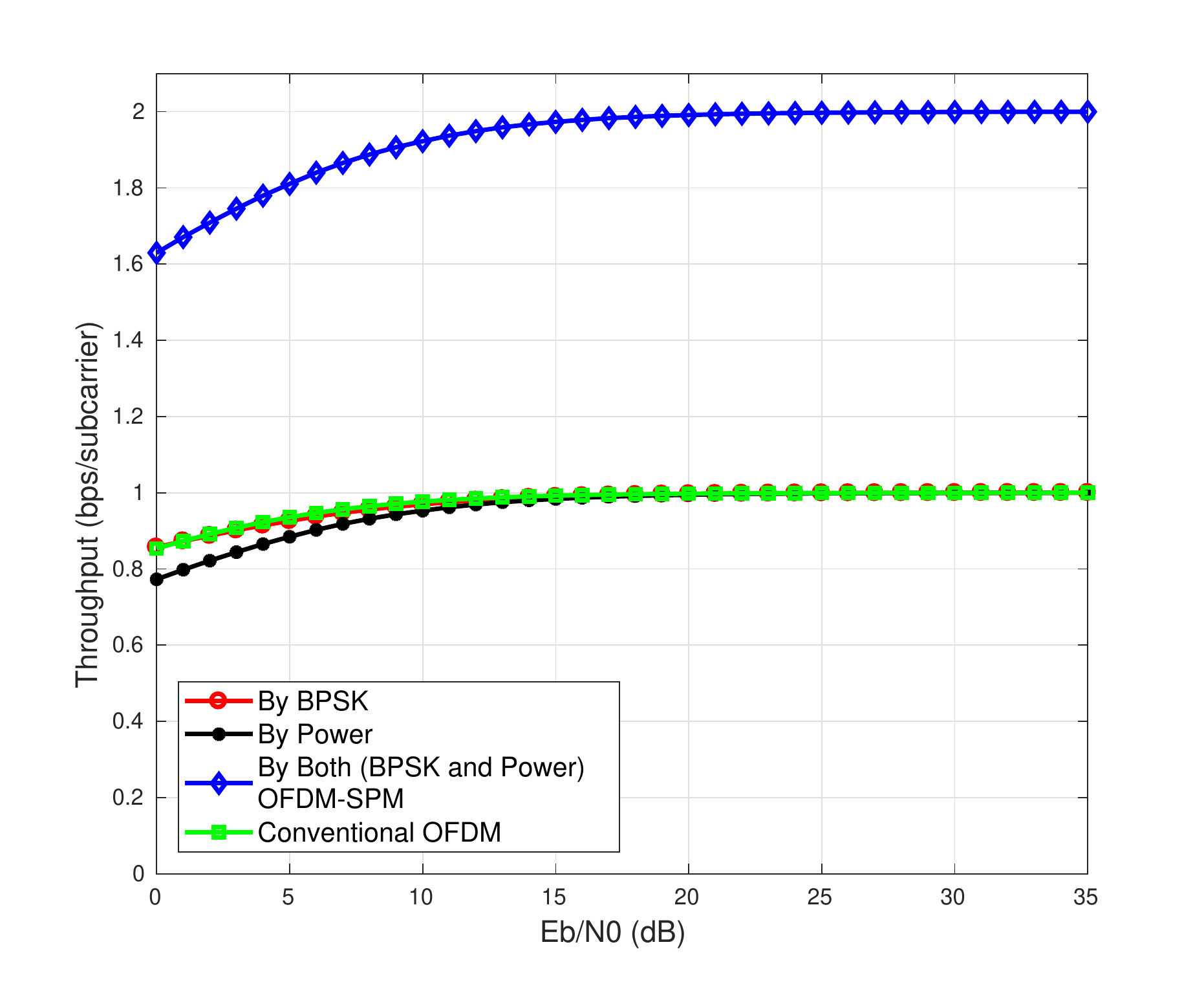}
\caption{Throughput of OFDM-SPM with optimized power reallocation policy, where the optimal power levels that minimize the overall error rates are used.}
\end{figure}

\section{Conclusion}
In this paper, a novel modulation technique capable of utilizing a third dimension (power) to convey extra data bits named as OFDM-SPM was introduced. OFDM-SPM manipulates the power of the subcarriers of an OFDM block and uses it as an extra degree of freedom to convey extra data bits. OFDM-SPM, when compared to conventional OFDM was shown to require only half the number of subcarriers required by conventional OFDM to transmit the same number of bits. This gives OFDM-SPM many advantages over conventional OFDM, as it doubles the spectral efficiency, saves power, reduces transmission delays and makes it capable of reducing complexity as well by using half the IFFT size that conventional OFDM uses. The saved power can either be saved or reallocated. These two power policies characterize OFDM-SPM with flexibility to accommodate different applications and use cases. If the power is saved, the scheme exhibits a certain degree of degradation in the BER in comparison to the BER of conventional OFDM with BPSK. However, by utilizing the saved power and reallocating it to the subcarriers of the OFDM symbol, an enhanced BER performance can be achieved while maintaining the other gains that OFDM-SPM delivers. It is notable to mention that other methods in conjunction with OFDM-SPM can be used to improve the BER performance such as using coding techniques and antenna receiver diversity schemes; however, this is left for future research works and studies.


The inherent working principles of OFDM-SPM allow it to evade various flaws exhibited in other schemes, where OFDM-SPM does not suffer from the error propagation seen in OFDM-SIM, and unlike OFDM-IM, its transceiver structure is of negligible complexity (i.e., no ML detection, no mapping tables, no dependent errors).
The complexity that OFDM-SPM adds to the system transceiver structure is minor and in fact is capable of reducing the system complexity further by reducing the IFFT size used, while achieving the same throughput that conventional OFDM can deliver. 

Besides, OFDM-SPM can be seen as a scheme capable of transmitting two separate data streams. By observing the BER curves of OFDM-SPM regardless of the power scheme followed, it is seen that the stream of bits carried by the power of the subcarriers is more erroneous than the bits carried by the BPSK symbols. This introduces the possibility of using OFDM-SPM to serve different users with different bit error rate requirements. However, the investigation of this research direction is left for future work.  


These aforementioned merits of OFDM-SPM without a doubt stress its significance and the gains it could add to beyond 5G communication systems (e.g., 6G). furthermore, it signifies that it should be additionally examined and researched from various perspectives, where the pairing of OFDM-SPM with other schemes can bring better performing schemes. Until now, further research on this scheme aims at examining OFDM-SPM and its performance when paired with higher modulation orders to attain even much higher data rates by utilizing both the in-phase and quadrature subcarrier components such as M-PSK. Being a novel scheme with great potential, it is of most extreme significance to really comprehend the magnitude of the benefits of OFDM-SPM, and of the advantages it could bring by utilizing it in future communication systems (6G and beyond).


%

\appendices
\textcolor{black}{
\section{The relation of the proposed OFDM-SPM design to 6G networks}} \textcolor{black}{
One of the main challenges the existing standardized 5G designs for certain use-cases (e.g., URLLC) suffers from is the fact that the subcarrier spacing parameter, which corresponds to certain numerology in the OFDM waveform, has to be increased and become large enough to obtain short symbol intervals that are capable of meeting the low latency transmission requirements of URLLC services. However, such large subcarrier spacing used by high-order numerologies will inevitably reduce the amount of data rate that can be delivered within a single time slot, aside from the fact that the signaling overhead would increase in the case of using short time slots, resulting in significant reduction in the data rate/throughput in an outrageous manner. This specific shortcoming can make current 5G designs incapable of fulfilling and satisfying the requirements of new emerging applications such as extended reality (XR) services (including augmented, mixed, and virtual reality (AR/MR/VR)), haptics (tactile Internet), real-time gaming, telemedicine, brain-computer interfaces, flying vehicles, and connected autonomous systems \cite{Saad6GVision}. This is so because such applications require to be simultaneously supported not only by low latency with good reliability, but also with high throughput through increasing the overall spectral efficiency per device  [24]–[27]. Motivated by this observation, in this paper, a novel, low-complexity, and low-latency modulation scheme, which is capable of adding a third data-carrying dimension to double the spectral efficiency per device, is developed and proposed for meeting the future requirements of 6G and beyond networks.}


\section{Explaining the Power Allocation and Reallocation Equations}

Eq. (1) and (2) are presented once again for the convenience of the reader \begin{equation} L^2 + H^2 = 2E_{b}\end{equation}\begin{equation} L^2 + H^2 = 4E_{b}\end{equation}
As can be seen from Eq. (15) or equivalently Eq. (1), the constraint upon which the power levels \textit{L} and \textit{H} are found, states that the power of these levels should sum up to two times the bit energy \textit{E\textsubscript{b}}. This is to ensure that the power factors \textit{L} and \textit{H} do not result in an average power per subcarrier exceeding that which is used by conventional OFDM with BPSK, where the power per subcarrier is equivalent to \textit{E\textsubscript{b}}. Eq. (1), for the sake of clarity, can be re-written as \begin{equation} \frac{L^2 + H^2}{2} = E_{b}\end{equation} From Eq. (17), we understand that the average power of the low power and high power subcarriers must be equal to the bit energy \textit{E\textsubscript{b}}, and given that the appearance of a '0' and '1' is equiprobable which is a valid assumption, then the above equation holds.

Furthermore, the same logic applies to Eq. (16) or equivalently Eq. (2). This equation puts a constraint on the \textit{L} and \textit{H} power levels, such that the power of these levels should sum up to four times the bit energy \textit{E\textsubscript{b}}, or equally Eq. (2) can be rewritten as follows \begin{equation} \frac{L^2 + H^2}{2} = 2E_{b}\end{equation} From Eq. (18), we understand that the average power of the low power and high power subcarriers must be equal to two times the bit energy \textit{E\textsubscript{b}}. This is a fair strategy since a subcarrier in OFDM-SPM is capable of carrying two bits, which in conventional OFDM with BPSK would require two subcarriers, thus amounting to two times the bit energy \textit{E\textsubscript{b}} and given that the appearance of a '0' and '1' is equiprobable which is a valid assumption, then the above equation holds as well.

\section{Finding the optimal \textit{L} and \textit{H} values} 
As was mentioned, the optimal values of \textit{L} and \textit{H} in Eq. (1) and Eq. (2) were found by an exhaustive process of trial and error, with small sized increments to ensure that all the possible values of \textit{L} and \textit{H} were spanned, and their resulting performances compared. Taking Eq. (1), as an example, we set \textit{H} to an initial value of 1.05, since the bit energy \textit{E\textsubscript{b}} or symbol energy is normalized to 1, then \textit{L} can be found as 0.947. The simulations are run with these values and the results are observed, after that the value of \textit{H} is incremented slightly by 0.01 and the corresponding value of \textit{L} is found, and the results are observed and recorded and so on so forth until the optimal values of \textit{L} and \textit{H} are found. The optimal values of Eq. (2) were found in the same manner.

\section{OFDM-SPM Theoretical BER Analysis}
\par By referring to Fig. 2, we can use the constellation points of OFDM-SPM to derive the analytical expression of the bit error rate of OFDM-SPM. By looking at Fig. 2, it is intuitive to assume the derivation of the BER expression of OFDM-SPM will be done in a manner similar to that of 4-PAM. However, the important difference being that the changes exhibited in the BER expressions of OFDM-SPM will be to simulate the effect of the different Euclidean distances between the symbols. 

\par Since the constellation points on the right and left halves of Fig. 2 are symmetric, we can simply find the probability of error of the points on either half of the constellation diagram and then multiply the collective result by a factor of 2. Arbitrarily, for demonstration purposes, we choose the right half of the constellation map. 

\par Our starting point is by observing that the probability of a BPSK symbol at a signal to noise ratio (SNR) $\frac{E_b}{N_0}$ being found in error due to the effects of noise and the Rayleigh channel response on it. This effect on the BER performance can be given by the following equation \cite{OFDM-SIS}:

\begin{equation}
BER = \frac{1}{2}\bigg{(} 1 - \sqrt{\frac{\frac{E_b}{N_0}}{1 + \frac{E_b}{N_0}}}\bigg{ )},
\end{equation}
where the Euclidean distance between the symbols is a function of ratio between the energy of the symbols and noise variance (i.e., $\frac{E_b}{N_0}$). This equation can be used as a basis for finding the bit error rates of the power detection process of the symbols of OFDM-SPM.

\par Generally, we have two cases of error in the right half of the constellation map, namely Fig. 2. Either that a low power symbol is detected as high power symbol, or that a high power symbol is detected as a low power symbol. 

Firstly, we consider the case of a low power symbol being detected as a high power symbol. This can occur in one of two ways, either the symbol is detected as a high power symbol with amplitude (\textit{H$\sqrt{E_b}$}), with a probability of occurrence which can be found similar to that of Eq. (15), with a small change to simulate the change in the Euclidian distance between the two symbols. The minimum distance for the symbol to be detected as a symbol of amplitude (\textit{H$\sqrt{E_b}$}) is $\frac{H - L}{2}$. As such, the probability of this occurring, let's call it \textit{E\textsubscript{1}}:

\begin{equation}
E_{1} = \frac{1}{2}\bigg{(} 1 - \sqrt{\frac{(\frac{H - L}{2})^2 \frac{E_b}{N_0}}{1 + (\frac{H - L}{2})^2 \frac{E_b}{N_0}}} \bigg{)}
\end{equation}

It is also possible that the low power symbol can be detected as a high power symbol with amplitude (\textit{-H$\sqrt{E_b}$}) which is less likely to occur but  still possible. The minimum distance for the low power symbol to be detected as such is $\frac{H + 3L}{2}$, which gives us a probability of occurrence \textit{E\textsubscript{2}}:

\begin{equation}
E_{2} =  \frac{1}{2}\bigg{(} 1 - \sqrt{\frac{(\frac{H + 3L}{2})^2 \frac{E_b}{N_0}}{1 + (\frac{H + 3L}{2})^2 \frac{E_b}{N_0}}} \bigg{)}
\end{equation}

Assuming an equal probability of occurrence of 0's and 1's, the probability of the occurrence of such a symbol is $\frac{1}{4}$, giving the total probability of error as:

\begin{equation}
    \frac{1}{4}(E_{1} + E_{2})
\end{equation}

Now, we come to the probability of a high power symbol being detected as a low power symbol. Starting from the symbol with amplitude (\textit{H$\sqrt{E_b}$}), we see that the power bit conveying the power of this symbol will be detected in error if the symbol is detected as a low power symbol in either half of the constellation plane; however, if the noise is high enough such that the symbol is detected as a symbol with amplitude (\textit{-H$\sqrt{E_b}$}), the power bit detected will not be in error as the power of the symbol remains high. We can take this into  account by subtracting the probability of this occurring from the probability of the symbol being detected as a low power symbol. The minimum distance required for the high power symbol to be detected as a low power symbol is given as $\frac{H - L}{2}$, thus we obtain the probability of this occurrence \textit{E\textsubscript{3}} as:

\begin{equation}
E_{3} = \frac{1}{2}\bigg{(} 1 - \sqrt{\frac{(\frac{H - L}{2})^2 \frac{E_b}{N_0}}{1 + (\frac{H - L}{2})^2 \frac{E_b}{N_0}}} \bigg{)},
\end{equation}
which is identical to Eq. (16). The minimum distance required for the high power symbol in the right half of the constellation plane to be detected as a high power symbol in the left half plane is $\frac{3H + L}{2}$. As such, the probability of this occurrence \textit{E\textsubscript{4}} is found as 

\begin{equation}
E_{4} = \frac{1}{2}\bigg{(} 1 - \sqrt{\frac{(\frac{3H + L}{2})^2 \frac{E_b}{N_0}}{1 + (\frac{3H + L}{2})^2 \frac{E_b}{N_0}}} \bigg{)}
\end{equation}

Similar to the previous case, the probabilities are multiplied by a factor of $\frac{1}{4}$, and we obtain the following:

\begin{equation}
    \frac{1}{4}(E_{3} - E_{4})
\end{equation}

The total probability of error is thus a sum of the terms in Eq. (22) and Eq. (25). Additionally, since we assumed symmetry between the symbols of the left half and right half  of the constellation plane we multiply each of the expressions in Eq. (22) and Eq. (25) by a factor of 2 to take into account the error rates of the symbols of the left half plane as well, and because the terms \textit{E\textsubscript{1}} and \textit{E\textsubscript{3}} are identical, they can be summed. Thus, we are then left with

\begin{equation}
    E_{1} + \frac{1}{2}E_{2} - \frac{1}{2}E_{4},
\end{equation}

which is identical to the expression given by Eq. (10).

\textcolor{black}{
\section{The reason why OFDM-SPM is compared to OFDM-BPSK (not OFDM-QPSK)}
QPSK can be considered as two independent BPSK, each of which uses only one carrier (sin or cosine) to convey data (i.e., a QPSK signal essentially combines two orthogonally modulated BPSK signals by simultaneously using two orthogonal carriers). On contrast, the proposed OFDM-SPM scheme uses only one carrier, which is similar to the case of BPSK. Accordingly, QPSK consumes twice the power/energy of BPSK. Also, QPSK requires more complex processing at the receiver as it depends on having strict synchronization between I and Q components, whereas OFDM-SPM with BPSK does not have such as issue. In addition to this, for OFDM-SPM with BPSK modulation, the symbol duration for each bit is same as the bit duration used with conventional OFDM with BPSK, but for QPSK, the symbol duration is twice the bit duration (symbol duration of QPSK is twice that of BPSK). Thus, using OFDM with QPSK requires more time to send and receive than using OFDM-SPM with BPSK. Due to all these reasons, it makes complete sense to compare OFDM-SPM with OFDM-BPSK rather than OFDM-QPSK.}

\textcolor{black}{
\section{The trade-off relationship between Throughput and BER in OFDM-SPM}
As is the case with most communication designs, the existence of an inherent trade-off between BER and throughput or spectral efficiency, which are key indicators or metrics used for measuring the performance of wireless communication systems, is inevitable and unavoidable. Indeed, the trade-off in the proposed OFDM-SPM totally depends on the way we design OFDM-SPM and the perspective we look from it. For instance, when we design OFDM-SPM to provide a user device with two data streams simultaneously with closely similar error probability (as shown in Fig. 9 and Fig. 10 in the manuscript), then OFDM-SPM does not have much advantage in terms of BER and shows a slight degradation when compared to the BER of conventional OFDM with BPSK, due to using a threshold-based, non-coherent detector for half of the total bits (i.e., the bits modulated by subcarrier power), which turns out to be an additional advantage due to providing low-complexity processing that is desirable by IoT devices. On the plus side, OFDM-SPM doubles the throughput compared to conventional OFDM. On the other hand, when we design OFDM-SPM to provide two data streams simultaneously but with unequal error probability (as shown in Fig. 7 and Fig. 8), the stream of data bits modulated by BPBK in the OFDM-SPM scheme with power reallocation policy shows advantage in terms of BER (around 2-3 dB) and presents some enhancement when compared to the BER of conventional OFDM with BPSK, whereas the other data stream encoded in the power of subcarriers in OFDM-SPM is in fact an additional gain for sending more data bits, but at the expense of some degradation in the BER (as shown in Fig. 7 and Fig. 8).  In general, OFDM-SPM usually opts to focus on doubling the throughput, and thus introduces a large gain in the throughput without the need for additional bandwidth, which is very critical and useful for the data-hungry applications that require more throughput without stringent requirements on the BER such as real-time video streaming.}


\ifCLASSOPTIONcaptionsoff
  \newpage
\fi



\bibliographystyle{IEEEtran}



%
\begin{IEEEbiography}[{\includegraphics[width=1in,height=1.25in,clip,keepaspectratio]{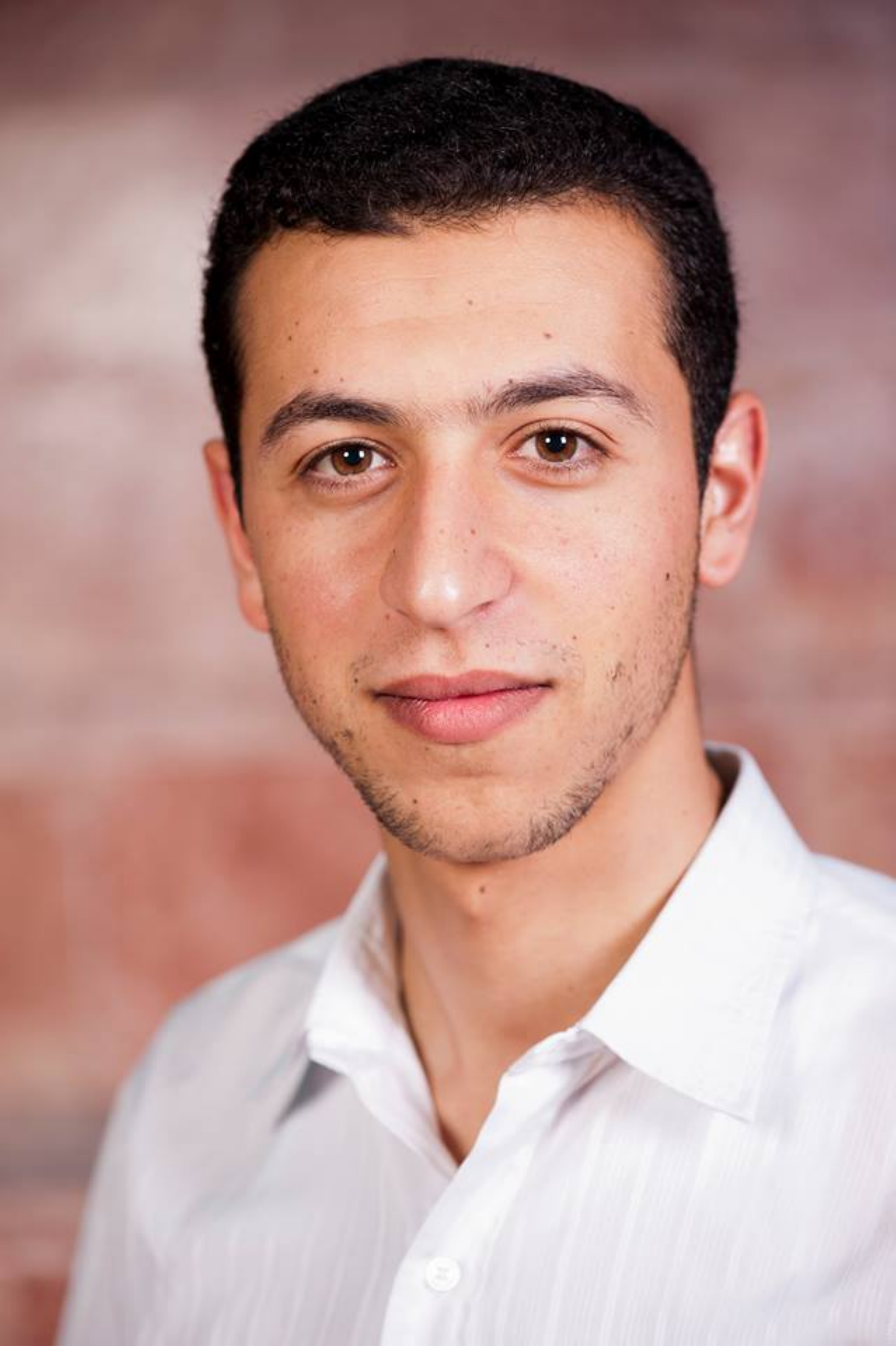}}]{Jehad M. Hamamreh}
received the B.Sc. degree in electrical and telecommunication engineering from An-Najah University, Nablus, in 2013, and the Ph.D. degree in electrical-electronics engineering and cyber systems from Istanbul Medipol University, Turkey, in 2018. He was a Researcher with the Department of Electrical and Computer Engineering, Texas A and M University. He is currently an Assistant Professor with the Electrical and Electronics Engineering Department, Antalya International (Bilim) University, Turkey. His current research interests include wireless physical and MAC layers security, orthogonal frequency-division multiplexing multiple-input multiple-output systems, advanced waveforms design, multi-dimensional modulation techniques, and orthogonal/non-orthogonal multiple access schemes for future wireless systems. He is a regular investigator and referee for various scientific journals as well as a TPC Member for several international conferences. He can be reached via e-mail: jehad.hamamreh@gmail.com // web: https://sites.google.com/view/wislab. 


\end{IEEEbiography}
\begin{IEEEbiography}[{\includegraphics[width=1in,height=1.25in,clip,keepaspectratio]{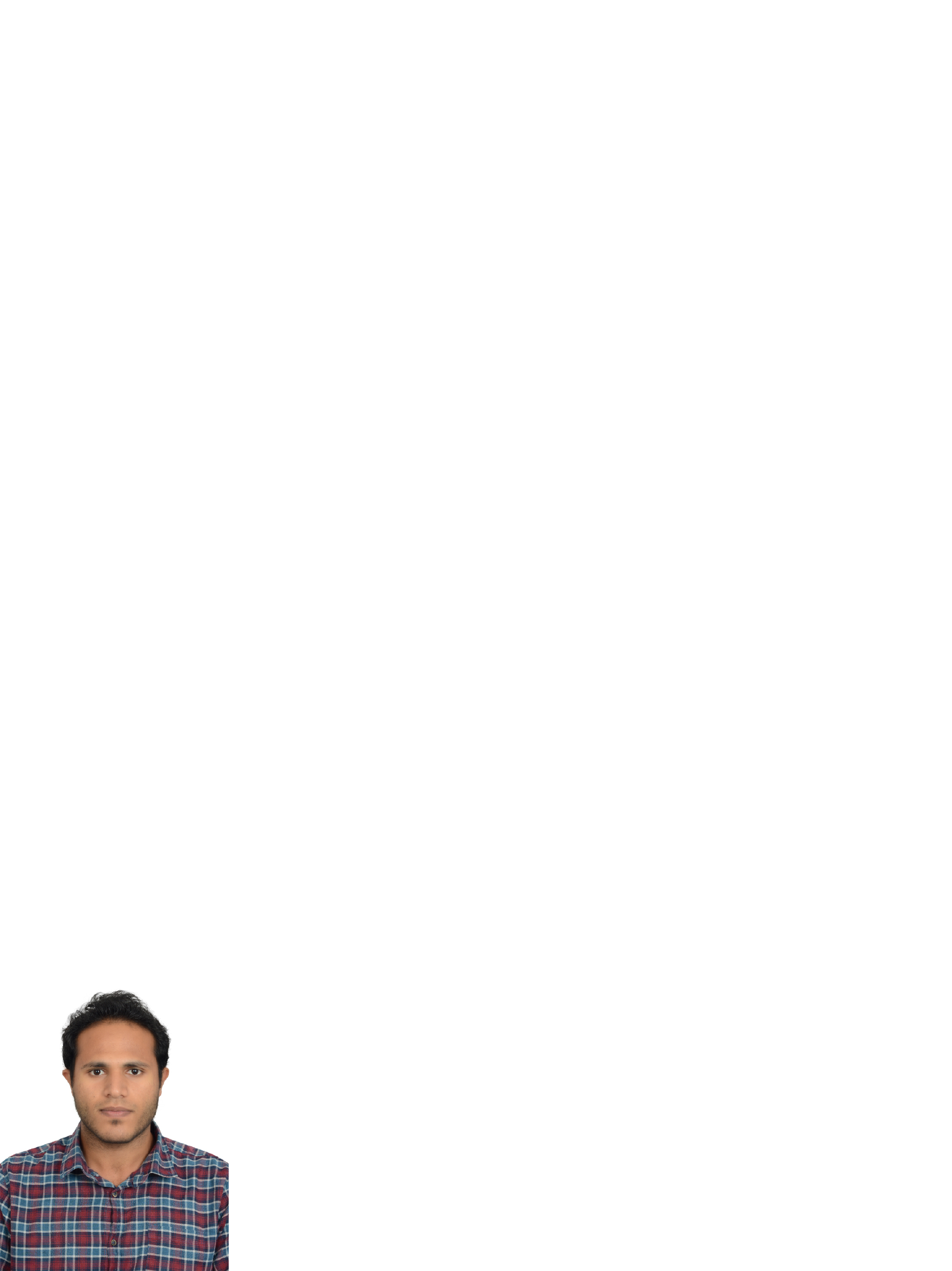}}]{Abdulwahab Hajar}
recently received his B.Sc from the department of electrical and electronics engineering, Antalya Bilim University in Antalya, Turkey. \newline He is the co-author of three publications, and has worked with professor Jehad M. Hamamreh on the topic of OFDM-SPM.

\end{IEEEbiography}


\begin{IEEEbiography}[{\includegraphics[width=1in,height=1.25in,clip,keepaspectratio]{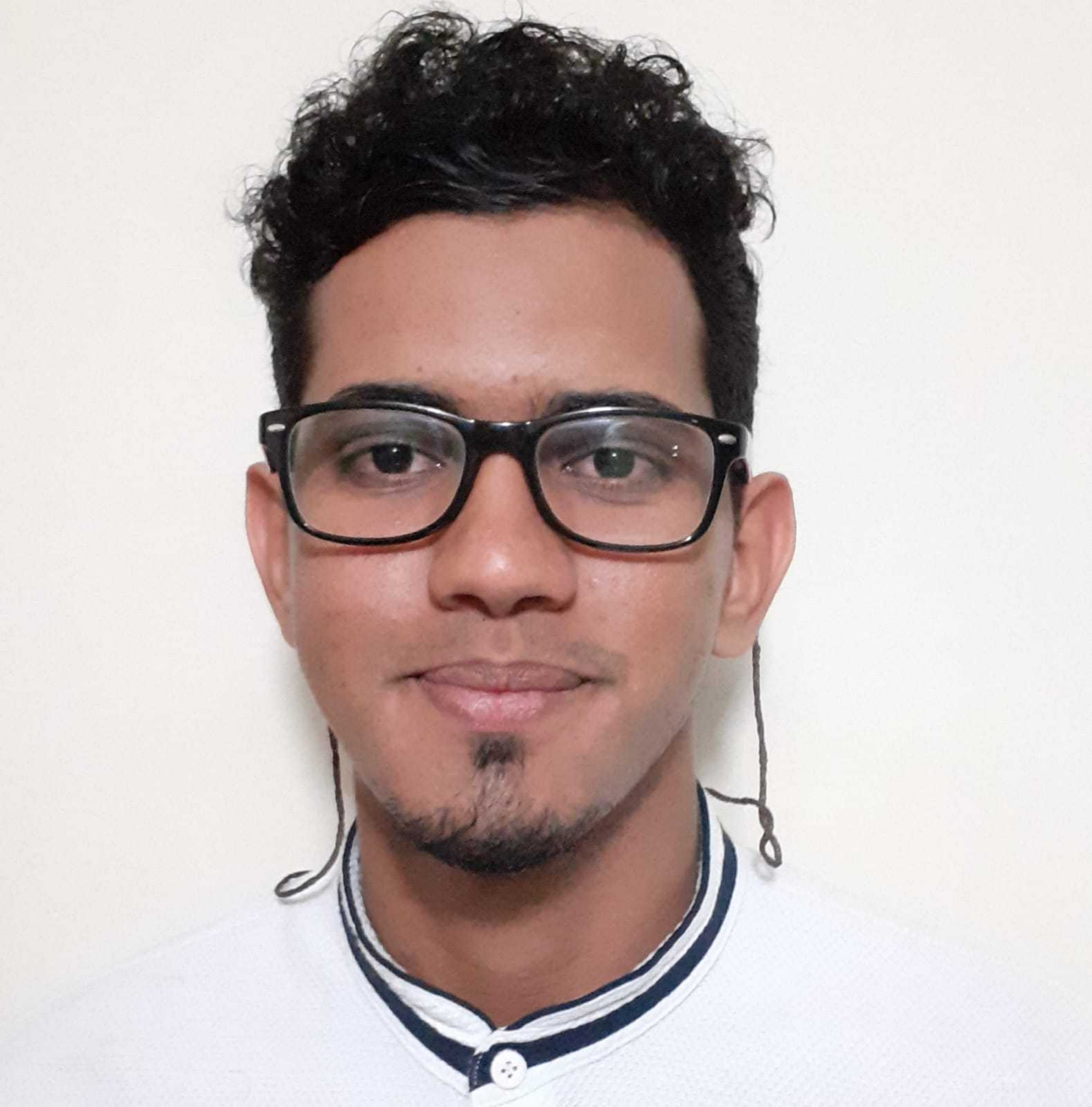}}]{Mohamedou Abewa}
recently received his B.Sc in electrical and electronics engineering from Antalya Bilim University. He is the co-author of three publications, and is currently working with professor Jehad M. Hamamreh on the topic of OFDM-SPM.
\end{IEEEbiography}
\end{document}